\documentclass[12pt]{iopart}

\usepackage{graphicx}  
\begin{document}

\title{Pulsing mechanism based on power adiabatic evolution of pump in Tm-doped fiber laser}

\author{Fuyong Wang}

\address{School of Information and Electrical Engineering, Hebei University of Engineering, Handan 056038, China}
\ead{jiaoyi@sjtu.edu.cn}
\vspace{10pt}

\begin{abstract}
We prove an alternative pulsing mechanism based on power adiabatic evolution of pump in Tm-doped fiber laser. A pulsed laser technique, unlike Q-switching and gain-switching, is explored under the equilibrium between the stimulated emission and absorption. After the laser reaching a CW steady state, the temporal fluctuation of pump power is called power adiabatic evolution if it does not break the balance between the stimulated emission and the absorption. Under the pump power adiabatic evolution the population densities in the upper and lower laser levels get clamped to their steady-state values, and the temporal shape of output laser is identical with that of pump. Based on the power adiabatic evolution of pump a laser pulse can therefore be generated by a pulsed pump which is added to a CW pump. Moreover, the temporal profile, duration and peak power of the laser pulse can all be precisely controlled. Power adiabatic evolution of pump can be achieved in the three pumping shemes of Tm-doped fiber laser, provided that the duration of pump pulse is broader than a certain value. The study on the pulsing mechanism highlights an alternative pulsed technique possessing several advantages over Q-switching and gain-switching. 
\end{abstract}

%
%
%
%
%

\section{Introduction}
Owing to an increasing need for laser pulses in widespread applications including medicine, micromachining and manufacturing, a huge amount of attention has been paid to pulsed laser techniques \cite{Shijie2015,Vid2014,Roka2013,Vida2013,Larsen2014,Yun2005,Isinsu2018,Swiderski2013,Swiderski2004,Yanming2004,Gong2008,Jan2011}. Pumping the population inversion to a value far in excess of the threshold population and then suddenly switching the cavity Q factor from a low to a high value allows the generation of laser pulse with a duration from a few nanoseconds to a few tens of nanoseconds. That is Q-switching technique which is widely used to realize pulsed operation \cite{Oleg2007,Shougui2018,Junqing2018,Wohlmuth2009}. To pump the inversion up to a very high value the stimulated emission is blocked by a high loss inside laser cavity and the population inversion keeps growing due to the pumping. After the population inversion has been built up to a high value far exceeding that of the threshold, the cavity loss is switched to a low value rapidly. Then the stimulated emission is dominant over the absorption and hence a laser pulse is generated. 

Rapidly switching the laser gain, instead of the cavity Q factor, to a high value can also allow a laser pulse generation. That is gain-switching technique, which is realized by exploiting a pump pulse that is so fast that the laser gain and hence the population inversion reach a value considerably above the threshold before a laser pulse has had time to build up to deplete the population inversion \cite{Orazio2010}. After the population inversion reaching a much larger value than the threshold value, a laser pulse occurs later in the tail of the pump pulse, which drives the inversion to a value well below the threshold. To achieve stable pulse generation, the laser gain must be switched off fast to prevent the inversion growing thereafter \cite{Min2007}. 

Q-switching and gain-switching are the two principle approaches for nanosecond pulse generation \cite{Orazio2010}. To produce a laser pulse both the Q-switching and gain-switching must undergo two processes: (1) Absorption process, where the population in the upper laser level keeps growing while the sitmulated emisson is absent; (2) Stimulated emission process, where the absorption is much weaker than stimulated eimission and the population inversion keeps decreasing. Those two techniques have the same pulsing mechanism, which is based on the nonequilibrium between the absorption and stimulated emission. Large fluctuation of population inversion deviating from the threshold value is a reflection of this nonequilibrium. However, this pulsing mechanism has some drawbacks and limitations. The temporal profile, duration and peak power of the laser pulse generated in both Q-switching and gain-switching cannot be precisely controlled.   

In our previous work, we have introduced a novel pulsing method in Yb-doped fiber laser, based on which a laser pulse with a tunable duration and controllable temporal profile can be generated \cite{Fuy2018,Fuyong2018}. We attributed the pulse-shaping in the approach to the power of seeded laser \cite{Fuyong2018}. The dynamics of the pulsing in this technique however does not be elaborated explicitly. Although some requirements on the pump pulse for employing this method have been pointed out \cite{Fuyong2018,Fuy2018}, the dependence of those requirements on the parameters and the requirement on the energy levels in the pulsed technique have not been investigated. Whether this pulsing method can be adopted in other fiber lasers, besides Yb-doped fiber laser, is unknown. Therefore, it is necessary to launch a deep investigation on the novel pulsed approach.   

In this paper, we continue to study the novel pulsed technique. We interpret the pulsing dynamics from the perspective of the balance between the stimulated emission and the absorption instead of from the seeded power, which provides a more profound understanding of the technique. In addition, we discuss the conditions of the pulsing technique in Tm-doped fiber laser, which has multiple absorption bands and promising applications \cite{Creeden2014,Yulong2013,Dickinson2000,Stuart1999,Mengmeng2015}. The study on the pulsing mechanism and on the conditions of the novel pulsed method highlights an alternative pulsing technique possessing several advantages over Q-switching and gain-switching. 

\section{Numerical model and rate equations of Tm-doped fiber laser}

\begin{figure*}[htbp]
\centerline{\includegraphics[width=\linewidth]{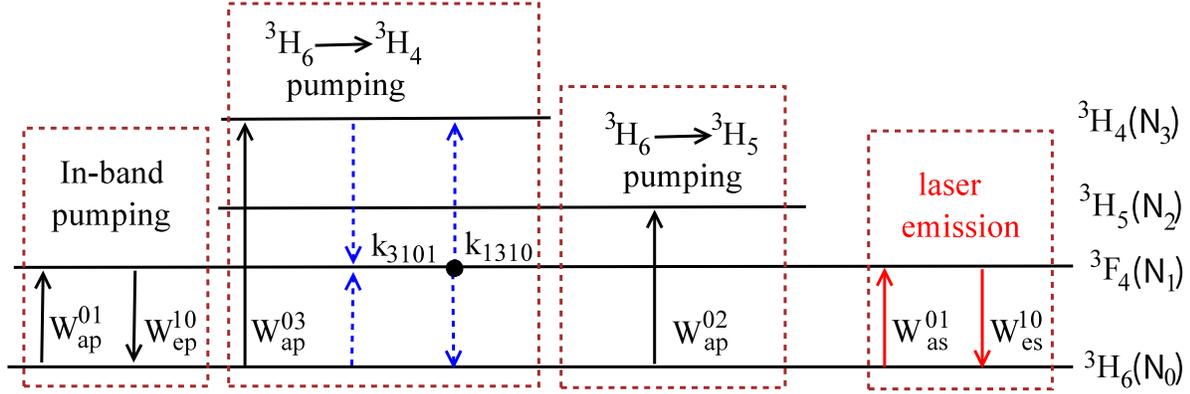}}
\caption{Simplified energy level diagram of Tm$^{3+}$ ions displaying. The pump and laser transitions are indicated as black and red solid arrows, respectively. The blue dash lines represent the cross-relaxation process in $^3H_6$ $\rightarrow$ $^3H_4$ pumping scheme.}\label{setup1}
\end{figure*}

Three different pumping schemes with the same laser emission are displayed in the simplified energy level diagram of Tm$^{3+}$, as seen in Fig. \ref{setup1}. The ground state pump absorption rates of those three pumping schemes are represented by W$_{ap}^{01}$, W$_{ap}^{02}$ and W$_{ap}^{03}$, respectively. The stimulated emission rate is W$_{es}^{10}$. For all of the pump schemes considered in the investigation, the governing equations of the forward (P$_{sf}$) and backward (P$_{sb}$) propagating laser power can be written as 
\begin{equation}
\frac{\partial P_{sf}}{\partial z}+\frac{1}{\upsilon _s}\frac{\partial P_{sf}}{\partial t}=\Gamma _s(\sigma_{es}^{10}N_1-\sigma_{as}^{01}N_0)P_{sf}-\alpha_sP_{sf}+ 2\sigma_{es}^{10}N_1\frac{hc^2}{\lambda _{s}^3}\Delta \lambda _s, \label{sf}
\end{equation}
\begin{equation}
-\frac{\partial P_{sb}}{\partial z}+\frac{1}{\upsilon _s}\frac{\partial P_{sb}}{\partial t}=\Gamma _s(\sigma_{es}^{10}N_1-\sigma_{as}^{01}N_0)P_{sb}-\alpha_sP_{sb}+ 2\sigma_{es}^{10}N_1\frac{hc^2}{\lambda _{s}^3}\Delta \lambda _s, \label{sb}
\end{equation}
where, laser absorption process is included with $\sigma _{as}^{01}$ representing the absorption cross section. The stimulated transition (with emission cross section of $\sigma _{es}^{10}$) from energy level $^3F_4$ (with population density of N$_1$) to energy level $^3H_6$ (with population density of N$_0$) in the thulium atoms results in an emission of laser with wavelength $\lambda _s$ around 2000 nm. $\Gamma _s$ is the overlap factor between the signal laser and the doped fiber area. A, c and h are the core area of fiber, the speed of light in a vacuum and the Planck constant, respectively. The group velocity of the signal laser propagating in fiber is represented by $\upsilon _s$. $\alpha _s$ represents the background loss of the fiber at the laser wavelengths. $\Delta \lambda _s$ is the bandwidth of the amplified spontaneous emission at around 2.0 $\mu$m.

The blue dash lines in Fig. \ref{setup1} represent the cross-relaxation process in $^3H_6$ $\rightarrow$ $^3H_4$ pumping scheme with cross-relaxation constants of k$_{3101}$ and k$_{1310}$. In the following, rate equations, which govern pump power and population densities of involved energy levels, are modeled separately in each of the three pump schemes.

\subsection{$^3H_6$ $\rightarrow$ $^3F_4$ Pump Scheme}

In the case of upper laser level $^3F_4$ being pumped directly, the population dynamics of two-level laser manifolds mainly depends on the stimulated absorption and emission of pump and signal. With stimulated emission rate of pump W$_{ep}^{10}$ and stimulated absorption rate of signal W$_{as}^{01}$ taken into account, the rate equations of population can be expressed as \cite{Stuart1999}
\begin{equation}
\frac{\partial N_1}{\partial t}=W_{ap}^{01}+W_{as}^{01}-\frac{N_1}{\tau_1}-(W_{es}^{10}+W_{ep}^{10}),\label{n1}
\end{equation}
\begin{equation}
N_0=N-N_1,\label{n2}
\end{equation}
where, $\tau_1$ is lifetime of $^3F_4$ level and N is the total doping concentration. The expressions of W$_{ap}^{01}$, W$_{ep}^{10}$, W$_{as}^{01}$ and W$_{es}^{10}$ are given by
\begin{equation}
W_{ap}^{01}=\frac{\lambda _p\Gamma _p}{hcA_{core}}\sigma_{ap}^{01}(\lambda _p)[P_{pf}+P_{pb}]N_0,\label{xa1}
\end{equation}
\begin{equation}
W_{ep}^{10}=\frac{\lambda _p\Gamma _p}{hcA_{core}}\sigma_{ep}^{10}(\lambda _p)[P_{pf}+P_{pb}]N_1,\label{xb2}
\end{equation}
\begin{equation}
W_{as}^{01}=\frac{\lambda _s\Gamma _s}{hcA_{core}}\sigma_{as}^{01}[P_{sf}+P_{sb}]N_0,\label{xc3}
\end{equation}
\begin{equation}
W_{es}^{10}=\frac{\lambda _s\Gamma _s}{hcA_{core}}\sigma_{es}^{10}[P_{sf}+P_{sb}]N_1,\label{xd4}
\end{equation}
where, $\lambda_p$ and $P_{pf,(pb)}$ are pump wavelength and forward (backward) power, respectively. A$_{core}$ is the area of fiber core and $\Gamma _p$ is the overlap factor between the pump and the doped fiber area.

As depleted mainly by absorption and increased due to deexcitation, the power of pump propagating along the Tm-doped fiber can be described by the following equations  
\begin{equation}
\frac{\partial P_{pf}}{\partial z}+\frac{1}{\upsilon_p}\frac{\partial P_{pf}}{\partial t}=\Gamma _p[\sigma_{ep}^{10}(\lambda _p)N_1-\sigma_{ap}^{01}(\lambda _p)N_0]P_{pf}-\alpha_pP_{pf},\label{p1}
\end{equation}
\begin{equation}
-\frac{\partial P_{pb}}{\partial z}+\frac{1}{\upsilon_p}\frac{\partial P_{pb}}{\partial t}=\Gamma _p[\sigma_{ep}^{10}(\lambda _p)N_1-\sigma_{ap}^{01}(\lambda _p)N_0]P_{pb}-\alpha_pP_{pb},\label{p2}
\end{equation}
where, $\upsilon_p$ and $\alpha _p$ are the group velocity of the pump propagating in fiber and the attenuation of pump, respectively.

\subsection{$^3H_6$ $\rightarrow$ $^3H_4$ Pump Scheme}

For the case where the population is pumped from $^3H_6$ to $^3H_4$ level, four engery levels are involved describing the  population dynamics. Since the lifetime of $^3H_5$ energy level is quite short ($\sim$ 7 ns) compared with other energy levels and most of the thulium ions at this level are rapidly transferred to $^3F_4$ level by nonradiative relaxation, the population density N$_2$ is negligible. With the cross-relaxation considered and the deexcitation of pump neglected, the rate equations governing the population and pump power are the following\cite{two1}
\begin{equation}
\frac{\partial N_3}{\partial t}=W_{ap}^{03}-(k_{3101}N_3N_0-k_{1310}N_1^2)-\frac{N_3}{\tau_3},\label{n3}
\end{equation}
\begin{equation}
\frac{\partial N_1}{\partial t}=W_{as}^{01}-W_{es}^{10}+\beta_{31}\frac{N_3}{\tau_3}+2(k_{3101}N_3N_0-k_{1310}N_1^2)-\frac{N_1}{\tau_1},\label{n4}
\end{equation}
\begin{equation}
N_0=N-(N_1+N_3),\label{n5}
\end{equation}
\begin{equation}
\frac{\partial P_{pf}}{\partial z}+\frac{1}{\upsilon _p}\frac{\partial P_{pf}}{\partial t}=-\Gamma _p\sigma_{ap}^{03}(\lambda _p)N_0P_{pf}-\alpha _pP_{pf},\label{p3}
\end{equation}
\begin{equation}
-\frac{\partial P_{pb}}{\partial z}+\frac{1}{\upsilon _p}\frac{\partial P_{pb}}{\partial t}=-\Gamma _p\sigma_{ap}^{03}(\lambda _p)N_0P_{pb}-\alpha _pP_{pb},\label{p4}
\end{equation}
where, $\tau_3$ is the fluorescence lifetime of $^3H_4$ level. $\beta_{31}$ is the branch ratio for the transition from level 3 to 1. The expression of W$_{ap}^{03}$ is given by
\begin{equation}
W_{ap}^{03}=\frac{\lambda _p\Gamma _p}{hcA_{core}}\sigma_{ap}^{03}(\lambda_p)[P_{pf}+P_{pb}].
\end{equation}
Here, $\sigma_{ap}^{03}$ is the stimulated absorption cross section. k$_{3101}$ and k$_{1310}$ are cross-relaxation coefficients of $^3H_4$ $\rightarrow$ $^3F_4$ $\&$ $^3H_6$ $\rightarrow$ $^3F_4$ process and $^3F_4$ $\rightarrow$ $^3H_4$ $\&$ $^3F_4$ $\rightarrow$ $^3H_6$ process, respectively.  

\subsection{$^3H_6$ $\rightarrow$ $^3H_5$ Pump Scheme}

For the case of pumping into the $^3$H$_5$ level of Tm$^{3+}$, the typical wavelength of pump is around 1.064 $\mu$m. To simplify the model we neglect pump induced excited state absorption and cross-relaxation processes, and then the rate equations describing the population densities in the three lowest energy levels of Tm$^{3+}$ are as follows
\begin{equation}
\frac{\partial N_2}{\partial t}=W_{ap}^{02}-\frac{N_2}{\tau_2},\label{n6}
\end{equation}
\begin{equation}
\frac{\partial N_1}{\partial t}=W_{as}^{01}-W_{es}^{10}-\frac{N_1}{\tau_1}+\frac{N_2}{\tau_2},\label{n7}
\end{equation}
\begin{equation}
N_0=N-(N_1+N_2),\label{n8}
\end{equation}
with the ground state pump absorption rate given by
\begin{eqnarray}
W_{ap}^{02}=\frac{\lambda_p\Gamma_p}{hcA_{core}}\sigma_{ap}^{02}(\lambda _p)[P_{pf}+P_{pb}]N_0.
\end{eqnarray}
The rate equations for the pump power propagating along the doped fiber are given by
\begin{equation}
\frac{\partial P_{pf}}{\partial z}+\frac{1}{\upsilon _p}\frac{\partial P_{pf}}{\partial t}=-\Gamma_p\sigma_{ap}^{02}(\lambda _p)N_0P_{pf}-\alpha_pP_{pf},\label{p5}
\end{equation}
\begin{equation}
-\frac{\partial P_{pb}}{\partial z}+\frac{1}{\upsilon _p}\frac{\partial P_{pb}}{\partial t}=-\Gamma_p\sigma_{ap}^{02}(\lambda _p)N_0P_{pb}-\alpha_pP_{pb},\label{p6}
\end{equation}

\subsection{The boundary conditions and parameters}

\begin{figure*}[htbp]
\centerline{\includegraphics[width=\linewidth]{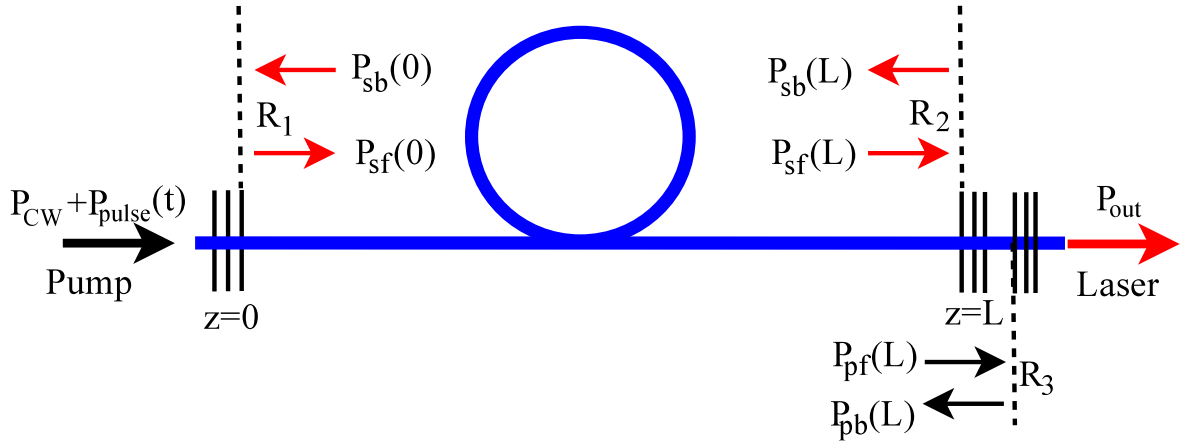}}
\caption{Schematic representation of the fiber laser geometry: P$_{sf,(sb)}$ and P$_{pf,(pb)}$ are the pump and laser powers respectively propagating in the forward (backward) z-direction.}\label{con}
\end{figure*}

A linear cavity consisting of a gain fiber with a length of L doped with Tm ions at a constant density N per unit volume is assumed in our simulation, as described schematically in Fig. \ref{con}. The laser power is reflected by Bragg reflector 1 (with reflectivity of R$_1$ and located at $z=0$) and Bragg reflector 2 (with reflectivity of R$_2$ and located at $z=L$). In this laser configuration, the boundary conditions of the laser power described by rate Eqs. (\ref{sf}) and (\ref{sb}) are expressed by 
\begin{equation}
P_{sf}(z=0,t)=R_1P_{sb}(z=0,t),
\end{equation}
\begin{equation}
P_{sb}(z=L,t)=R_2P_{sf}(z=L,t),
\end{equation}
\begin{equation}
P_{out}(z=L,t)=(1-R_2)P_{sf}(z=L,t),
\end{equation}
where P$_{out}$ is the output laser power. 

The pump power supplied by a continuous wave (CW) pump (with power denoted by P$_{CW}$) and a pulsed pump (with power of P$_{pulse}$(t)) is injected into the doped fiber at one side of cavity (at $z=0$). Generally, a fraction of pump power will be reflected back into the fiber at $z=L$. In order to include the pump power reflection, we assume another Bragg reflector (with reflectivity of R$_3$) for the pump wavelength is placed behind of the reflector 2 at the output end of the fiber laser. Since the corrugated sections of the Bragg reflector is much shorter than the fiber length, the pump reflector is assumed to be located at $z=L$. Thus, the boundary conditions of pump power described by rate Eqs. (\ref{p1}), (\ref{p2}), (\ref{p3}), (\ref{p4}), (\ref{p5}) and (\ref{p6}) are as follows
\begin{equation}
P_{pf}(z=0,t)=P_{CW}+P_{pulse}(t),
\end{equation}
\begin{equation}
P_{pb}(z=L,t)=R_3P_{pf}(z=L,t).
\end{equation}

\begin{table}[htbp]
\newcommand{\tabincell}[2]{\begin{tabular}{@{}#1@{}}#2\end{tabular}}
  \caption{\label{t1}Main parameters in the simulations \cite{two1,Stuart1998,Stuart1999,Mengmeng2015,Baofu2016}}
  \begin{center}
    \begin{tabular}{ccccccccc}
    \hline
     \tabincell{c}{Parameter}& Value& Parameter& Value  \\
    \hline
    $\tau_1$           &   334.7 $\mu$s               &   $\sigma_{as}^{01}$           &      0.1$\times 10^{-25}$ m$^2$ \\
    $\tau_2$           &     7 ns                     &   $\sigma_{es}^{10}$           &      5.0$\times 10^{-25}$ m$^2$ \\
    $\tau_3$           &   14.2 $\mu$s                &   $\sigma_{ap}^{03}$(793 nm)   &      5.0$\times 10^{-25}$ m$^2$ \\
$\sigma_{ep}^{10}$(1550 nm)  & 0.1$\times 10^{-25}$ m$^2$ & $\sigma_{ap}^{01}$(1550 nm)&      1.5$\times 10^{-25}$ m$^2$ \\
$\sigma_{ep}^{10}$(1900 nm)  & 5.0$\times 10^{-25}$ m$^2$ & $\sigma_{ap}^{01}$(1900 nm)&      0.25$\times 10^{-25}$ m$^2$\\
    $\alpha_s$         & 2.3$\times 10^{-3}$ m$^{-1}$ &   $\sigma_{ap}^{02}$(1064 nm)  &      1.2$\times 10^{-26}$ m$^2$ \\
    $\alpha_p$(793 nm) & 1.2$\times 10^{-2}$ m$^{-1}$ &   $\alpha_p$(1900 nm)          &      1.15$\times 10^{-2}$ m$^{-1}$ \\
    $\alpha_p$(1550 nm)& 1.07$\times 10^{-3}$ m$^{-1}$&   $\Delta \lambda_s$           &      300 nm     \\
    $k_{3101}$         & 3.0$\times 10^{-23}$ $m^3/s$ &   $k_{1310}$                   &      2.4$\times 10^{-24}$ $m^3/s$ \\  
    $R_1$              &     0.99                     &   $R_3$                        &      0.04 \\    
    \hline
    \end{tabular}
  \end{center}
\end{table}

Those three different pump schemes illustrated in Fig. \ref{setup1} require different pumping wavelengths. With the wavelength of
laser fixed at 2000 nm, 793 nm and 1064 nm are chosen as pumping wavelength in $^3H_6$ $\rightarrow$ $^3H_4$ and $^3H_6$ $\rightarrow$ $^3H_5$ pump scheme, respectively. As for the in-band pump scheme ($^3H_6$ $\rightarrow$ $^3F_4$), two typical wavelengths centering at 1550 nm and 1900 nm are used as those of pump. 

In $^3H_6$ $\rightarrow$ $^3F_4$ and $^3H_6$ $\rightarrow$ $^3H_5$ pump schemes core-pumped is adopted. The Tm-doped fiber has a core$/$cladding diameter of 10$/$130 $\mu$m with core numerical aperture of 0.15 and a population density of 8.3$\times 10^{25}$ $m^{-3}$ in those two pump schemes. In core-pumped configruation, the overlapping factors of pump ($\Gamma _p$) and signal ($\Gamma _s$) can be approximated by singlemode propagation in the core \cite{Marcuse1978}. In $^3H_6$ $\rightarrow$ $^3H_4$ pump scheme a Tm-doped fiber with a core$/$cladding diameter of 25$/$250 $\mu$m, a core numerical aperture of 0.21 and a population density of 2.85$\times 10^{26}$ $m^{-3}$ is chosen as the gain fiber. In the case of light propagating in the cladding, the overlapping factor of pump ($\Gamma _p$) is approximated by $\frac{A_{core}}{A_{clad}}$, where $A_{core}$ and $A_{clad}$ are the core and clad area, respectively. The main parameters used in our simulation are listed in Table \ref{t1}.

\section{Simulations and discussion}

In the simulation a fourth-order Runge-Kutta method is applied to solve Eqs. (\ref{n1}),(\ref{n2}),(\ref{n3}),(\ref{n4}),(\ref{n5}),(\ref{n6}),(\ref{n7}),(\ref{n8}) describing the dynamics of the population densities, while a modified Lax-Friedrichs scheme is used to solve Eqs. [(\ref{sf}),(\ref{sb})],[(\ref{p1}),(\ref{p2})],[(\ref{p3}),(\ref{p4})],[(\ref{p5}),(\ref{p6})] governing the laser and pump powers. It is a hybrid modified Lax-Friedrichs Runge-Kutta algorithm that provides a more accurate description of the dynamics of the Fabry-Perot Tm-doped fiber lasers \cite{Haroldo2016}.

We initially simulate the Fabry-Perot Tm-doped fiber laser illustrated in Fig. \ref{con} pumped by a CW power which is above the threshold. A steady state of output laser can be reached under this CW pump power P$_{CW}$ if the simulation time is long enough. Then, we focus our attention on investigating the performance of output laser under a power upward fluctuation of pump. In the investigation, at time t=0 the laser already reaches a steady state pumped by P$_{CW}$. After time t=0, a pulsed pump at the same wavelength as that of P$_{CW}$ is turned on to generate a smooth pulse with power of P$_{pulse}$(t). In the following, the temporal characteristics of the output power from Tm-doped fiber laser pumped by P$_{CW}$ and P$_{pulse}$(t) is simulated and analysed.   

\subsection{Laser pulsing based on power adiabatic evolution of pump in $^3H_6$ $\rightarrow$ $^3F_4$ pump scheme}
\begin{figure*}[htbp]
\centerline{
\includegraphics[width=0.5\linewidth]{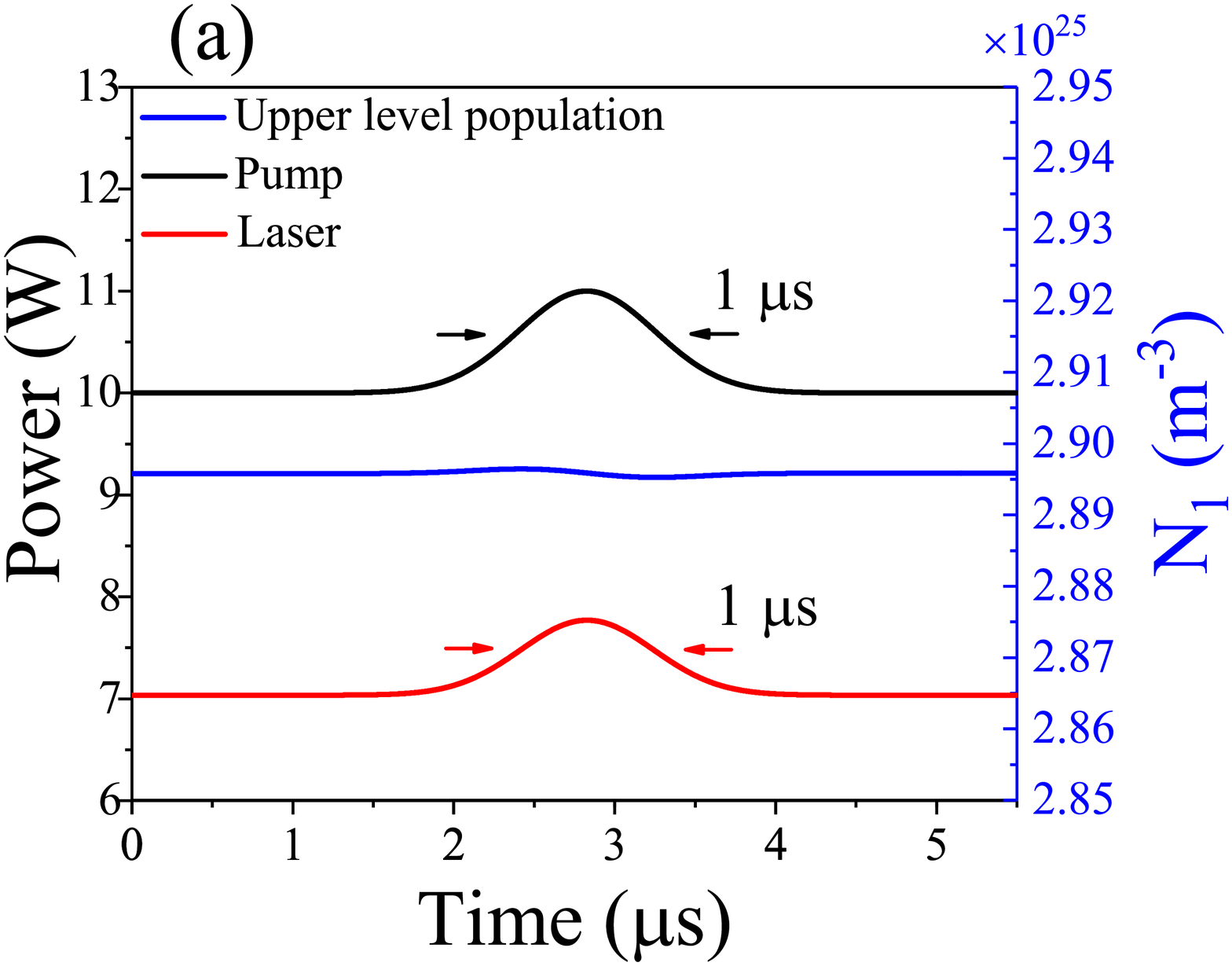}
\includegraphics[width=0.5\linewidth]{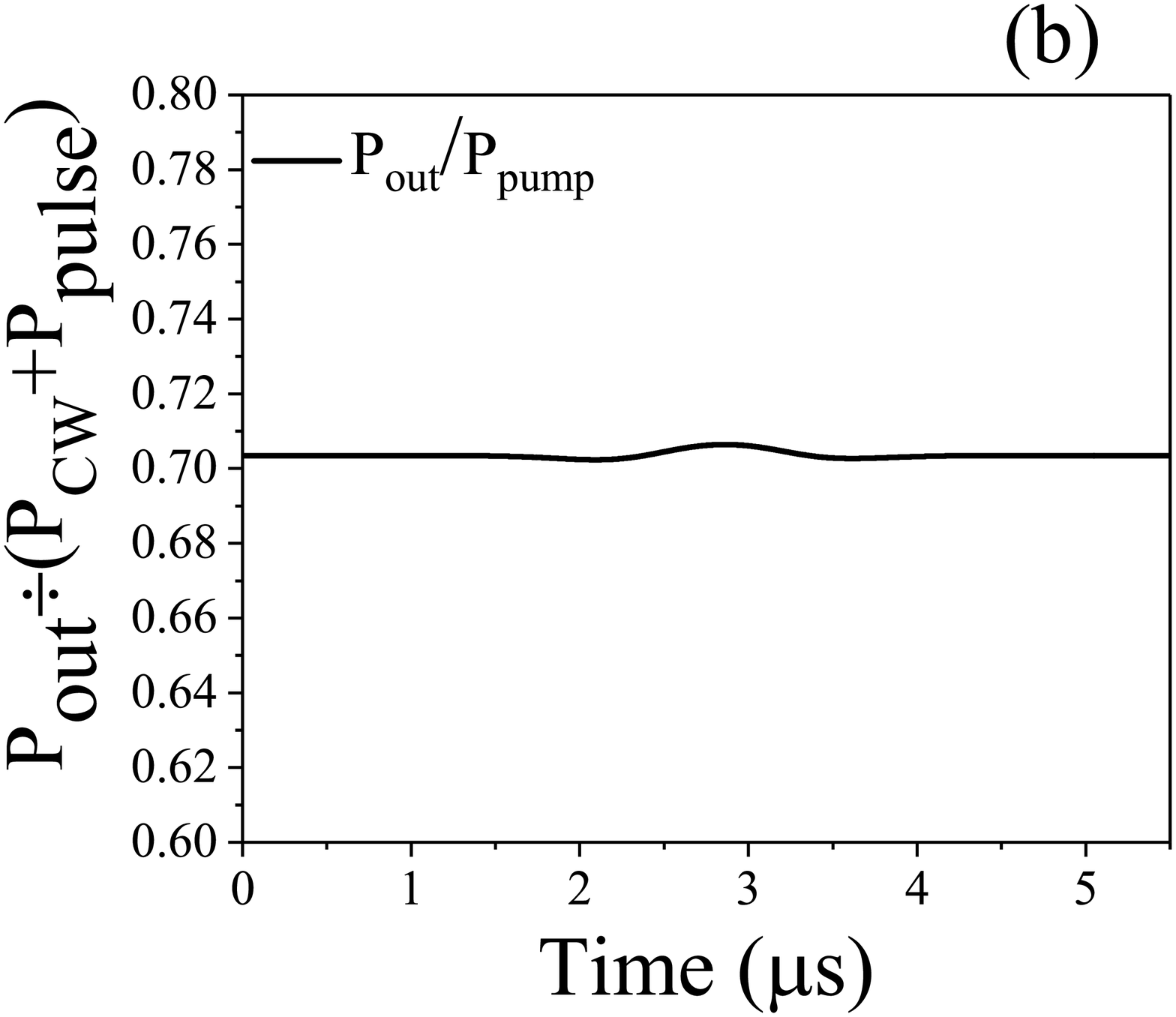}}
\caption{(a) The temporal evolution of the output power with that of pump and N$_1$(z=0,t) (at the position of z=0), and (b) the ratio of the output power to the input pump power when a Gaussion pump pulse P$_{pulse}$(t) with a duration of 1 $\mu$s and a peak power of 1 W is added to a P$_{CW}$ of 10 W, where P$_{pump}$=P$_{CW}$+P$_{pulse}$(t). These results are simulated under the active fiber length of 0.3 m, pumping wavelength of 1550 nm and R$_2$=0.1.}\label{ss}
\end{figure*}

In $^3H_6$ $\rightarrow$ $^3F_4$ pump scheme, the upper laser level ($^3F_4$) is pumped directly. Pumped by a 10 W CW power P$_{CW}$, a steady state of output laser is reached, and hence the stimulated emission and absorption are balanced. Starting from time t=0, a Gaussion pump pulse P$_{pulse}$(t) at 1550 nm (with duration of 1 $\mu$s and peak power of 1 W) is added to P$_{CW}$, creating a power upward fluctuation of pump. Under the pump power fluctuation, a smooth laser pulse is generated with both temporal profile and duration ($\sim $1 $\mu$s) nearly the same as those of pump pulse, as depicted in Fig. \ref{ss}(a). In addition, the population density N$_1$ of the upper laser level still gets clamped to the constant value under the pump power fluctuation (shown in Fig. \ref{ss}(a)). That means $\frac{\partial N_1}{\partial t}\approx 0$ and then Eq. (\ref{n1}) can be reduced to 
\begin{equation}
W_{ap}^{01}+W_{as}^{01}=\frac{N_1}{\tau_1}+(W_{es}^{10}+W_{ep}^{10}).\label{r1}
\end{equation}
According to Eqs. (\ref{n2}), (\ref{xa1}),(\ref{xb2}),(\ref{xc3}) and (\ref{xd4}), we get
\begin{equation}
\frac{N_1}{N}=\frac{\frac{\lambda _p\Gamma _p}{hcA_{core}}\sigma_{ap}^{01}P_p+\frac{\lambda _s\Gamma _s}{hcA_{core}}\sigma_{as}^{01}P_s}{\frac{\lambda _p\Gamma _p}{hcA_{core}}(\sigma_{ap}^{01}+\sigma_{ep}^{10})P_p+\frac{\lambda _s\Gamma _s}{hcA_{core}}(\sigma_{as}^{01}+\sigma_{es}^{10})P_s+\frac{1}{\tau _1}},\label{r2}
\end{equation}
where $P_s$ and $P_p$ represent $[P_{sf}+P_{sb}]$ and $[P_{pf}+P_{pb}]$, respectively. Since $\frac{1}{\tau _1}$ is much smaller than $\frac{\lambda _p\Gamma _p}{hcA_{core}}(\sigma_{ap}^{01}+\sigma_{ep}^{10})P_p+\frac{\lambda _s\Gamma _s}{hcA_{core}}(\sigma_{as}^{01}+\sigma_{es}^{10})P_s$ in our case, we ignore $\frac{1}{\tau _1}$ in the denominator of Eq. (\ref{r2}) and then get
\begin{equation}
\frac{N_1}{N}=\frac{\frac{\lambda _p\Gamma _p}{hcA_{core}}\sigma_{ap}^{01}+\frac{\lambda _s\Gamma _s}{hcA_{core}}\sigma_{as}^{01}\frac{P_s}{P_p}}{\frac{\lambda _p\Gamma _p}{hcA_{core}}(\sigma_{ap}^{01}+\sigma_{ep}^{10})+\frac{\lambda _s\Gamma _s}{hcA_{core}}(\sigma_{as}^{01}+\sigma_{es}^{10})\frac{P_s}{P_p}},\label{r3}
\end{equation}
which implies N$_1$ is a constant vaule on condition that $\frac{P_s}{P_p}$ is constant. The simulation result in Fig. \ref{ss}(b) indicates the ratio of the output laser power to the input pump power $\frac{P_{out}}{P_{CW}+P_{pulse}(t)}$ is nearly a constant. It implies that the stimulated emission and absorption processes still compensate one another under the pump power fluctuation. That is to say, the equilibrium between the stimulated emission and absorption in the case where a CW steady state of laser is reached under P$_{CW}$ is still maintained when P$_{pulse}$(t) is added. We consider this kind of temporal evolution of pump power as an adiabatic evolution, which does not break the equilibrium between the stimulated emission and absorption and can create a laser pulse generation with the same temporal profile and duration as those of pump.
\begin{figure*}[htbp]
\centerline{
\includegraphics[width=0.5\linewidth]{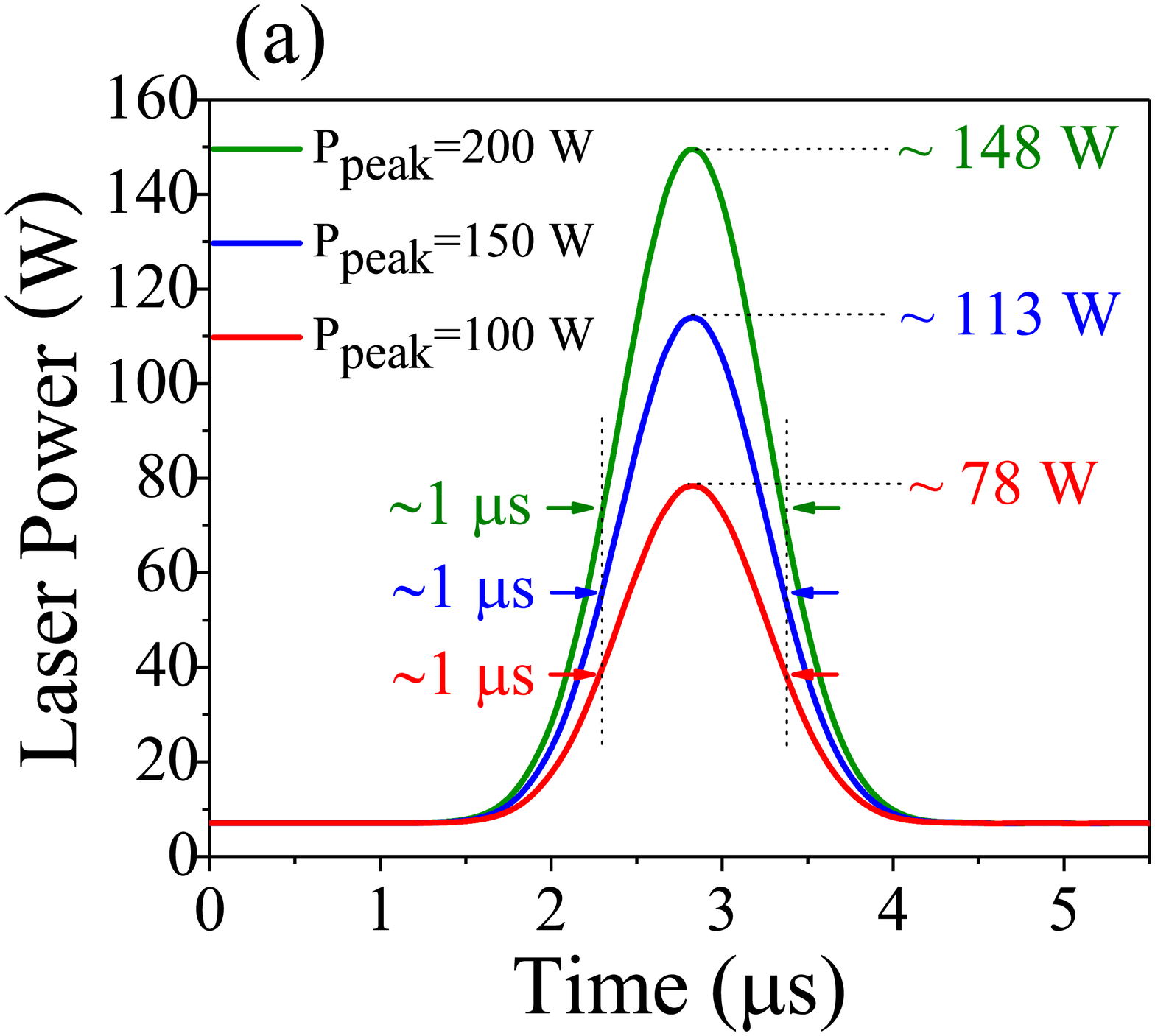}
\includegraphics[width=0.5\linewidth]{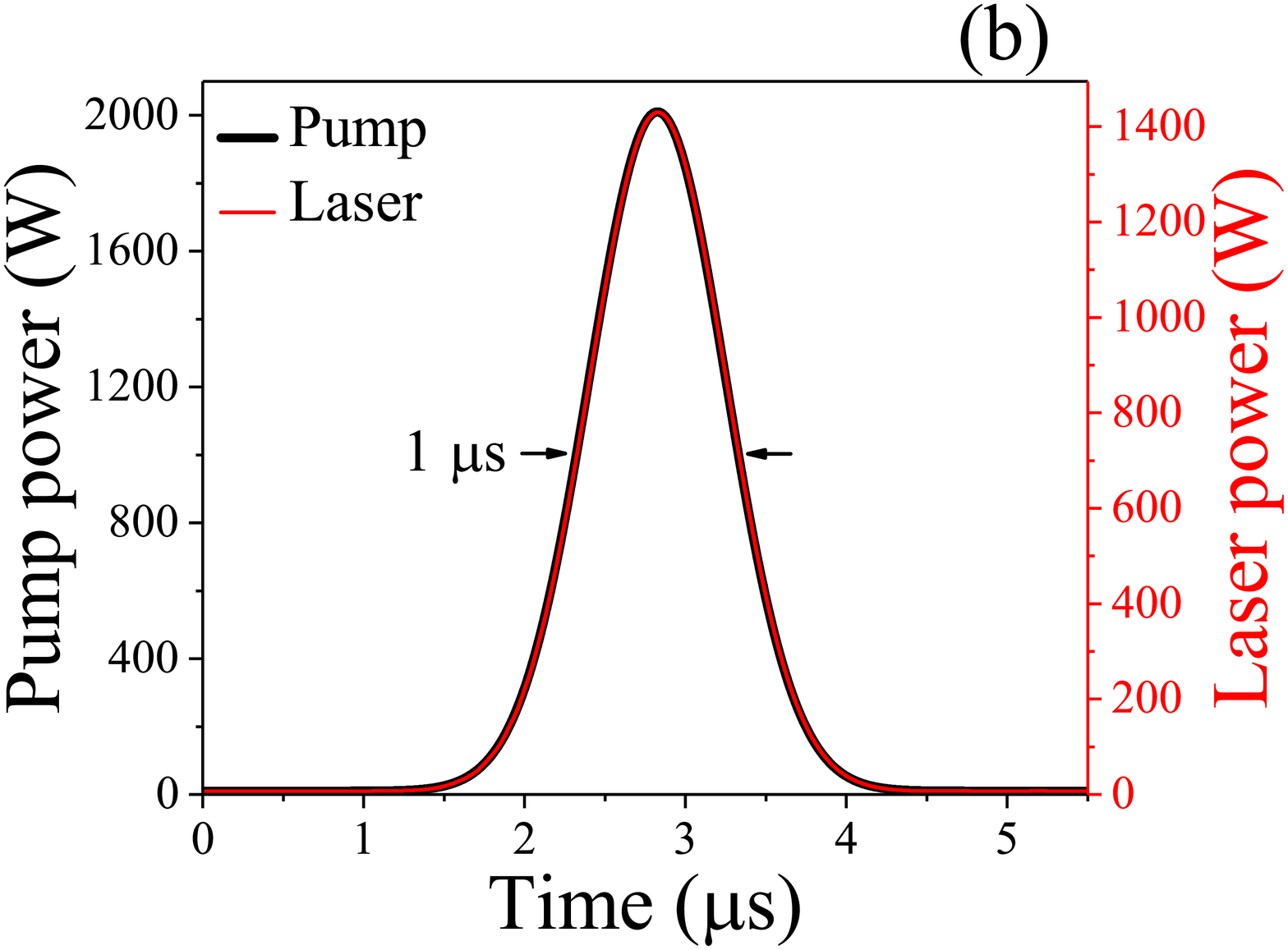}}
\caption{(a) The temporal characteristics of output laser pulses when the peak power of P$_{pulse}$(t) is increased to 100 W, 150 W, 200 W and (b) 2000 W. The peak power of P$_{pulse}$(t) is represented by $P_{peak}$. These results are simulated under the active fiber length of 0.3 m, pumping wavelength of 1550 nm and R$_2$=0.1.}\label{ss1}
\end{figure*}

When the power fluctuation of pump is enhanced by increasing the peak power of P$_{pulse}$(t) (denoted by P$_{peak}$) with the duration and temporal profile unchanged, the simulation results are shown in Fig. \ref{ss1}(a) and (b). The peak power of laser pulse generated by the pump power fluctuation increases with P$_{peak}$, and the ratio of laser peak power to pump peak power (P$_{CW}$+P$_{peak}$) approximately remains unchanged with P$_{peak}$ increasing, as displayed in Fig. \ref{ss1}(a). Even though P$_{peak}$ is 2000 W the duration and temporal profile of laser pulse are the same as those of pump exhibited in Fig. \ref{ss1}(b). The simulation results reveal that the absorption and the stimulated emission are in balance under the pump power fluctuation regardless of its peak power. Consequently, the power evolutions of pump in Fig. \ref{ss}(a), Fig. \ref{ss1}(a) and (b) are all adiabatic in the sense of the equilibrium between the stimulated emission and absorption. 

\begin{figure*}[htbp]
\centerline{
\includegraphics[width=0.5\linewidth]{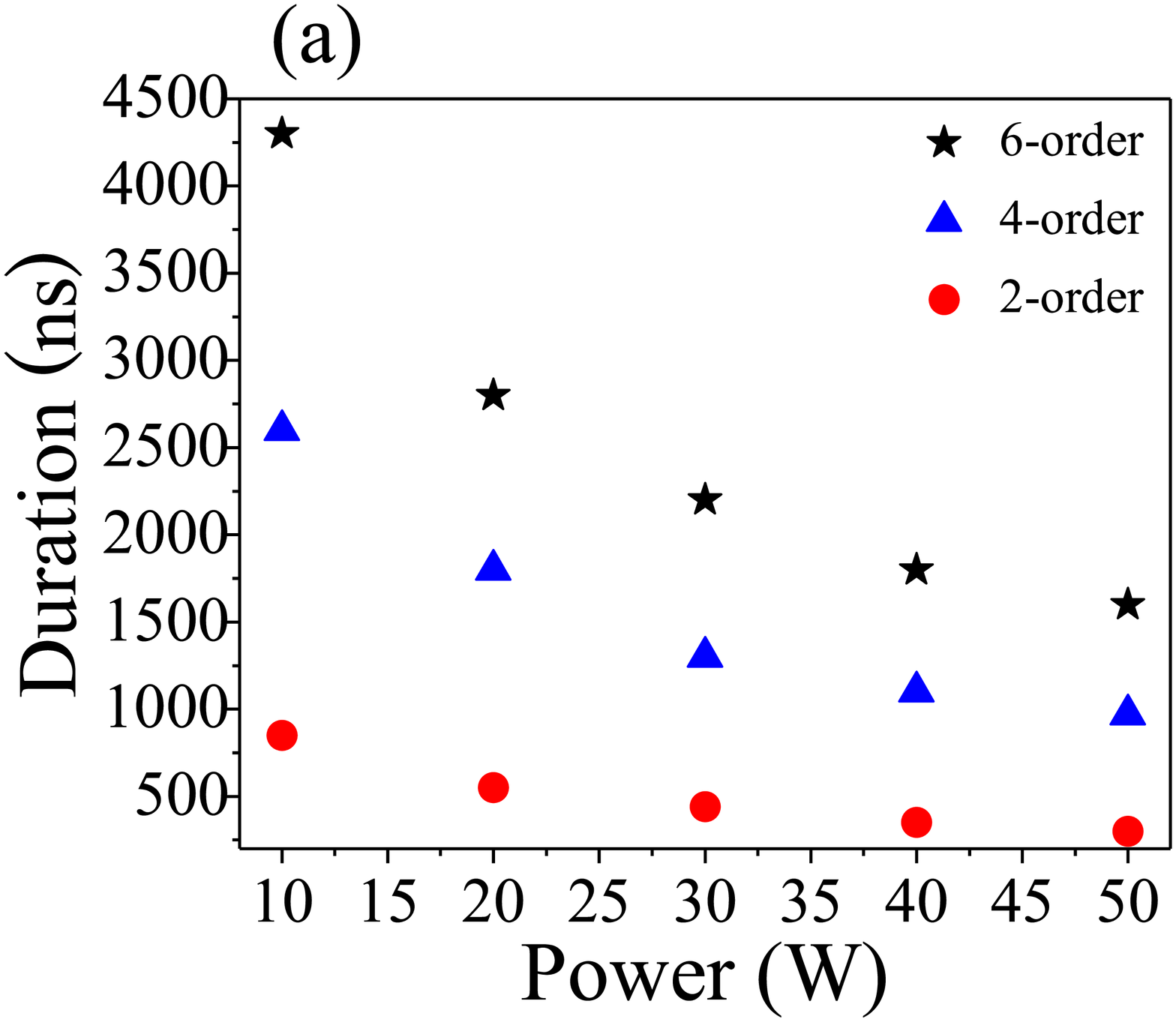}
\includegraphics[width=0.5\linewidth]{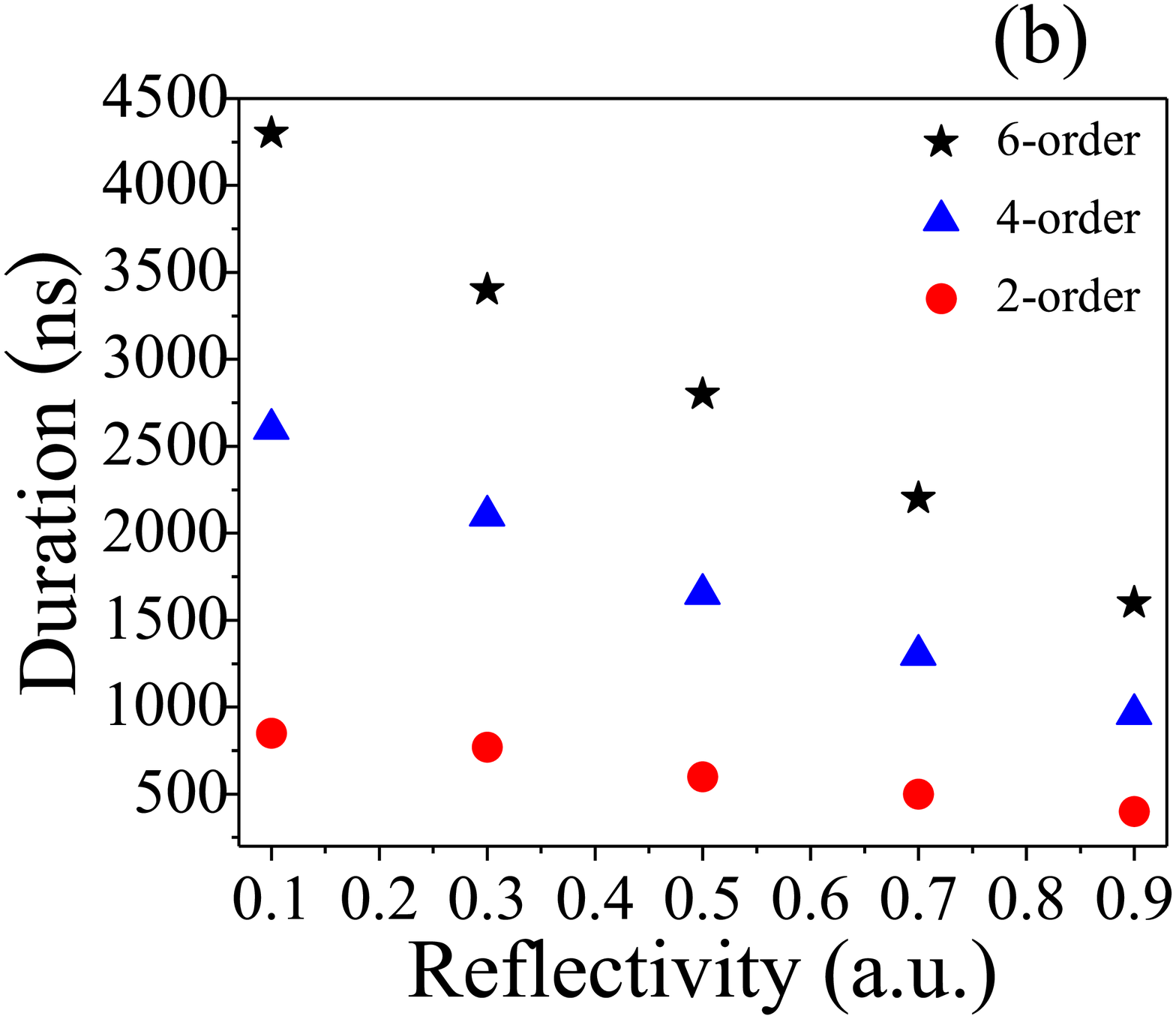}}
\caption{Dependence of the threshold duration $\tau_{th}$ of the pump pulse with a Gaussion (red circle), 4-order super-Gaussion (blue triangle) and 6-order super-Gaussion (black star) profile on (a) the CW pump power P$_{CW}$, and (b) the reflectivity of R$_2$. These results are simulated under the active fiber length of 0.3 m, pumping wavelength of 1550 nm and R$_2$=0.1.}\label{st}
\end{figure*} 

We have shown that the temporal profile of output laser is no longer the same as that of pump when the pump duration is shorter than a certain value \cite{Fuyong2018,Fuy2018}. It means that the power adiabatic evolution of pump requires the pump duration be broader than a corresponding certain value for each temporal profile. In order to facilitate the investigation, the pump duration at this value is named threshold duration of pump denoted by $\tau_{th}$. Then, we study the dependence of $\tau_{th}$ on the CW pump power P$_{CW}$ and the reflectivity of R$_2$.

The variations of threshold duration with P$_{CW}$ and R$_2$ are demonstrated in Fig. \ref{st}(a) and (b), respectively.
It can be observed that $\tau_{th}$ shortens with increasing P$_{CW}$ or R$_2$, as illustrated in Fig. \ref{st}. It is attributed to the fact that increasing P$_{CW}$ or R$_2$ can improve the power of laser inside the cavity which plays an important role in the laser pulse-shaping \cite{Fuyong2018}. In addition, the simulation results in Fig. \ref{st} show the threshold durations under pump pulses with three different temporal profiles: Gaussian profile (2-order), 4-order super-Gaussian profile, 6-order super-Gaussian profile. The threshold duration of 4-order super-Gaussian profile case is broader than that of Gaussian profile case and shorter than that of 6-order super-Gaussian profile case when the other parameters are the same as each other, shown in Fig. \ref{st}. It clearly indicates that the sharper the temporal profile edge of pump, the wider threshold duration is. Therefore, the CW pump power, the reflectivity of cavity reflector and the temporal profile can all influence the threshold duration greatly. And both increasing P$_{CW}$ (or R$_2$) and reducing the steepness of the temporal profile edges of pump pulse can support a more short threshold duration in pump power adiabatic evolution.      

\begin{figure*}[htbp]
\centerline{
\includegraphics[width=0.5\linewidth]{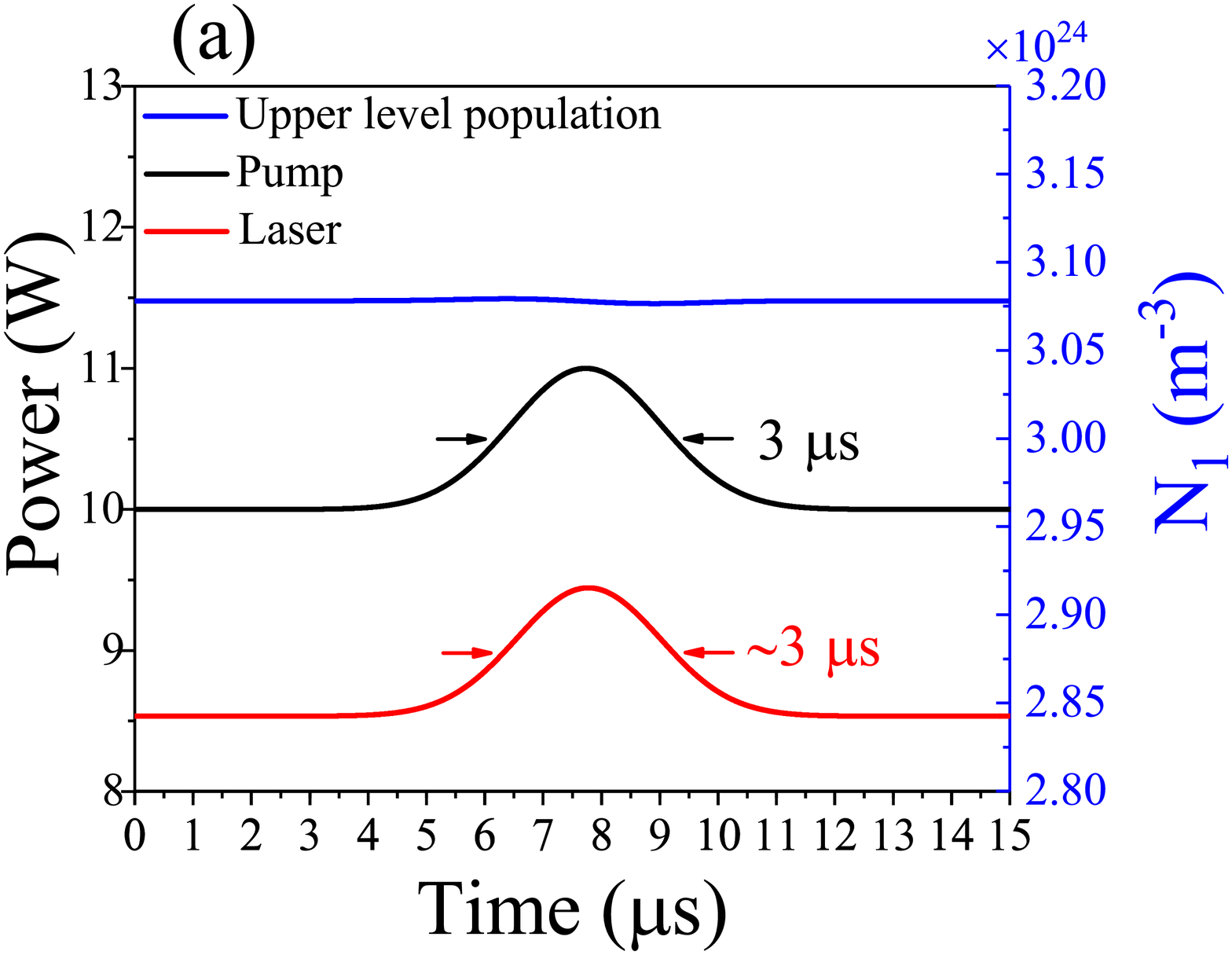}
\includegraphics[width=0.5\linewidth]{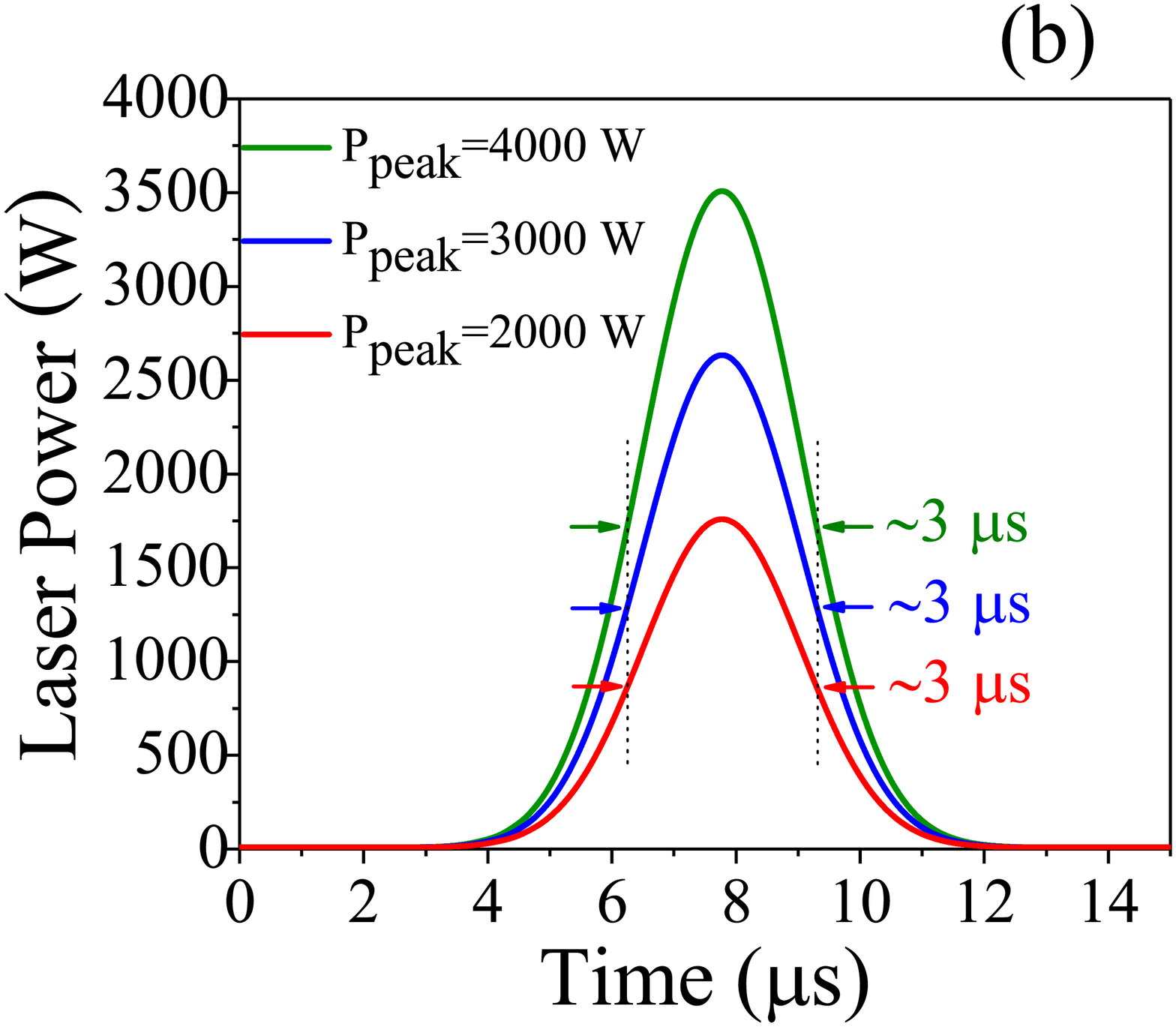}}
\caption{(a) The temporal evolution of the output power with that of pump and N$_1$(z=0,t) (at the position of z=0) when a Gaussion pulsed pump with a duration of 3 $\mu$s and a peak power of 1 W is added to a 10 W CW pump power. (b) The temporal characteristics of output pulses when the peak power of P$_{pulse}$(t) is increased to 2000 W, 3000 W, 4000 W. The peak power of P$_{pulse}$(t) is represented by $P_{peak}$. These results are simulated under the active fiber length of 4 m, pumping wavelength of 1900 nm and R$_2$=0.1.}\label{sp}
\end{figure*}

If the pumping wavelength is changed from 1550 nm to 1900 nm, the temporal evolutions of output laser power with that of pump are illustrated in Fig. \ref{sp}. A smooth and slow power upward fluctuation of pump (at 1900 nm) does not break the balance between the stimulated emission and absorption, as manifested in Fig. \ref{sp}(a). The temporal profile and duration of output laser are still the same as those of pump when the power fluctuation of pump is enhanced by increasing its peak power (as seem in Fig. \ref{sp}(b)). Therefore, the simulation results indicate that power adiabatic evolution can be effectively realized in the in-band pump scheme of Tm-doped fiber laser. 

According to the investigation above, a steady state of laser can tolerate a certain power upward fluctuation of pump without breaking the equilibrium between the stimulated emission and absorption in the in-band pumping scheme of Tm-doped fiber laser. In this case the pump power fluctuation with time is called power adiabatic evolution. Under power adiabatic evolution of pump the population in the laser upper and lower levels get clamped to their steady-state values and the power of output laser evolves in the same way as the pump, which creates a laser power fluctuation with the same temporal shape as that of pump. The power adiabatic evolution of pump requires the duration of the pump pulse be broader than a certain value which is named as threshold duration $\tau_{th}$. The threshold duration depends on the system parameters including the temporal profile of pump pulse, the reflectivity of cavity and the CW pump power. Fortunately, the power adiabatic evolution can still be maintained with an increasement of the peak power of pump. And the peak power of output laser increases with that of pump, while the temporal profile and duration of the laser are unchanged, as demonstrated in Fig. \ref{ss1}(a), (b) and Fig. \ref{sp}(b). Consequently, a laser pulse with tunable duration, controllable temporal shape and peak power can be generated based on the power adiabatic evolution of pump. In the next step, we attempt to explore whether the power adiabatic evolution of pump can be realized in other pumping schemes of Tm-doped fiber laser.

\subsection{Simulation results in $^3H_6$ $\rightarrow$ $^3H_4$ pump scheme}

\begin{figure*}[htbp]
\centerline{\includegraphics[width=0.9\linewidth]{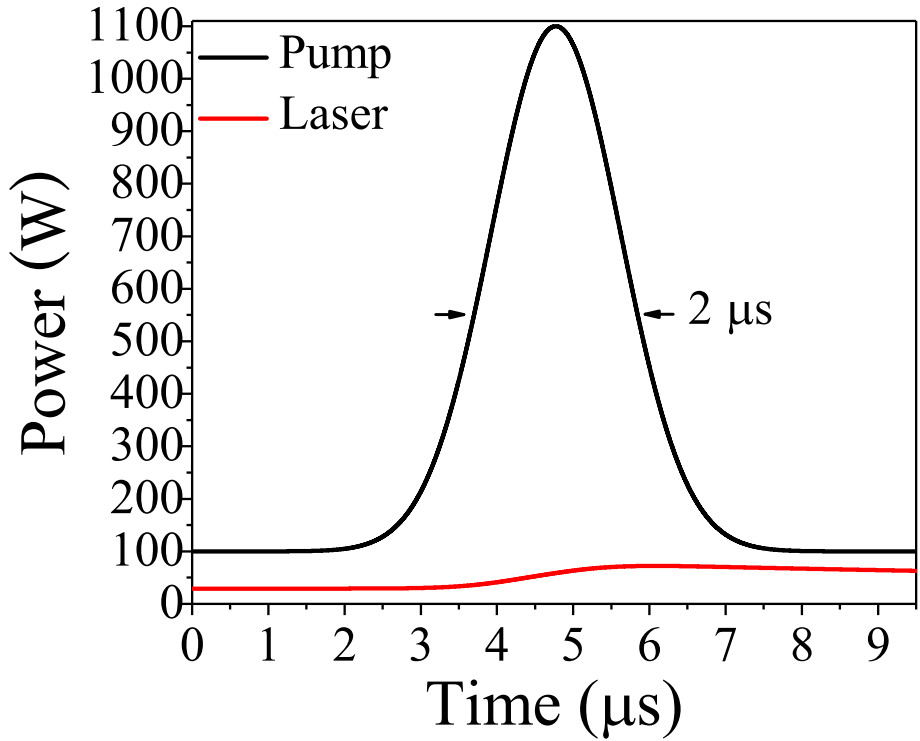}}
\caption{The temporal evolution of the output power with that of pump when a Gaussion pulsed pump with a duration of 2 $\mu$s and a peak power P$_{peak}$ of 1000 W is added to a P$_{CW}$ of 100 W. These results are simulated under the active fiber length of 2 m, pumping wavelength of 793 nm and R$_2$=0.5.}\label{sg}
\end{figure*}

In $^3H_6$ $\rightarrow$ $^3H_4$ pump scheme, a CW steady state of laser is reached pumped by a CW power (P$_{CW}$=100 W) at 793 nm, and then a Gaussion pump pulse (with duration of 2 $\mu$s and peak power of 1000 W) is added to the CW pump. The Change of output power brought about by the pump power fluctuation is exhibited in Fig. \ref{sg}. The temporal variation of laser power is much slower than that of pump, as is apparent in Fig. \ref{sg}. That is because the relaxation time ($\sim $14.2 $\mu$s) of population from $^3H_4$ to $^3F_4$ is longer than the duration of pump pulse (2 $\mu$s). The result in Fig. \ref{sg} demonstrates that the pump power adiabatic evolution cannot be realized when the duration of pump pulse is shorter than the relaxation time of population from pump level to upper laser level.  

\begin{figure*}[htbp]
\centerline{
\includegraphics[width=0.5\linewidth]{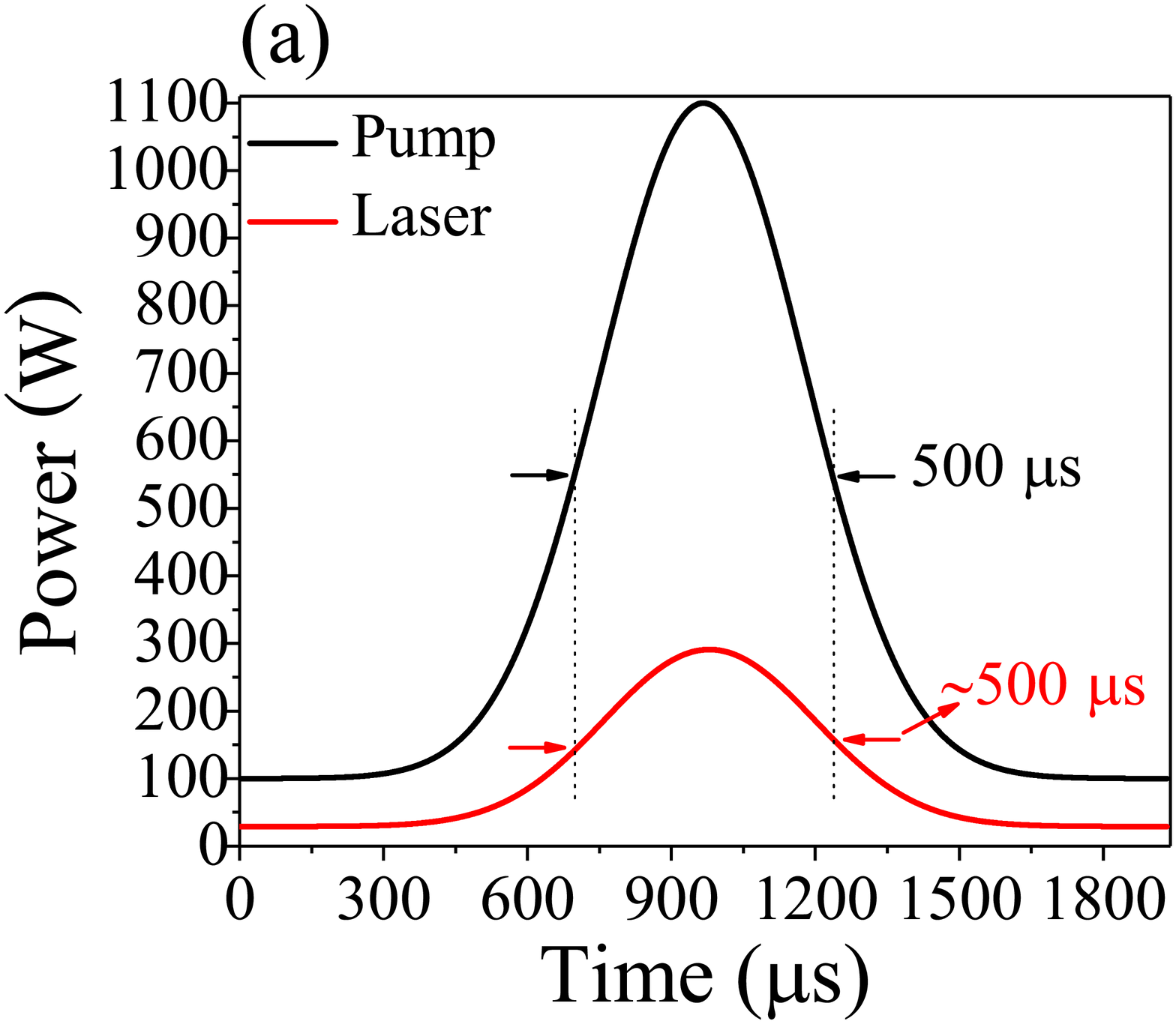}
\includegraphics[width=0.5\linewidth]{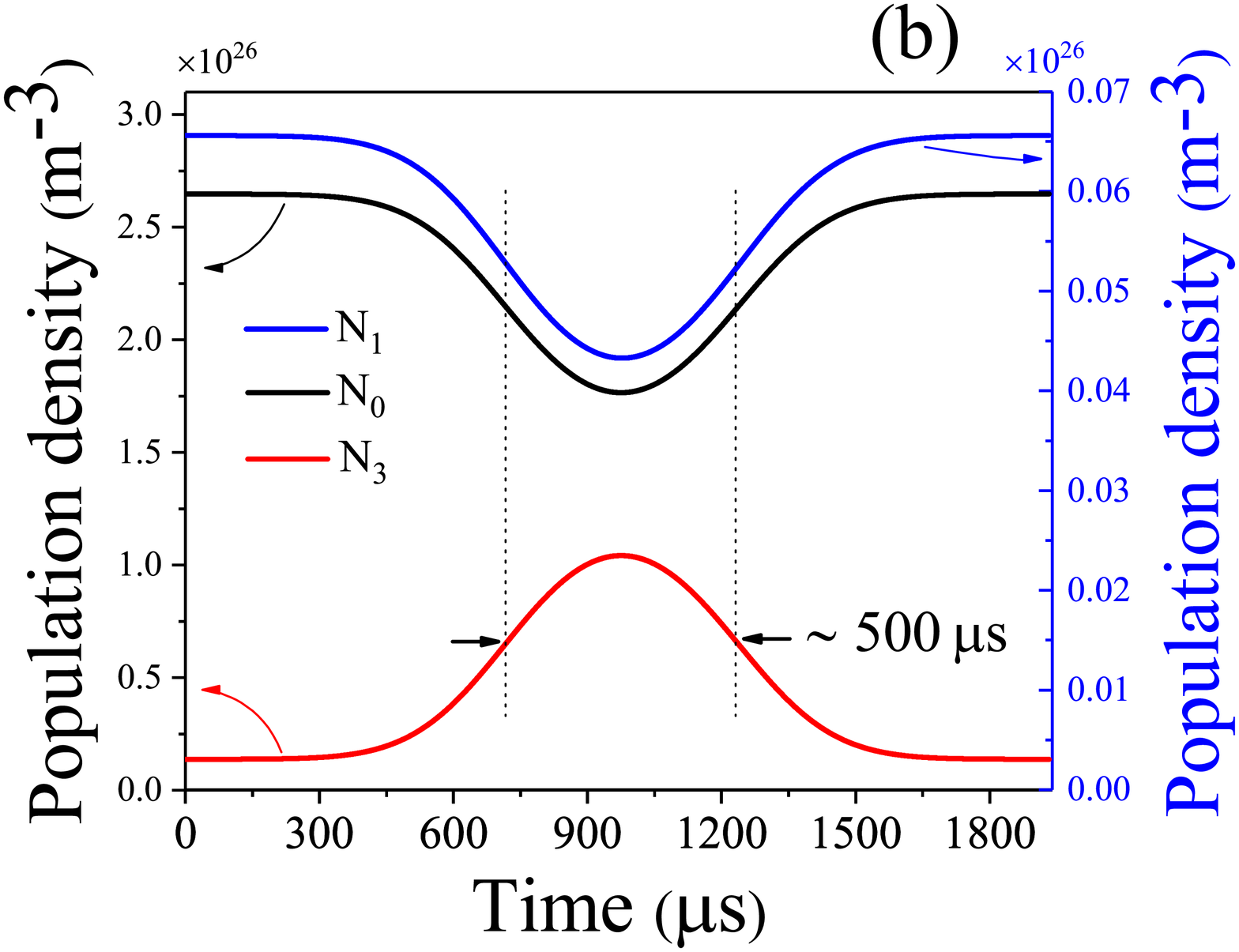}}
\caption{(a) The temporal evolution of the output power with that of pump, and (b) the temporal behaviors of N$_0$(z=0,t) in the lower laser level, N$_1$(z=0,t) in the upper laser level and N$_3$(z=0,t) in the pump level when a Gaussion pulsed pump with a duration of 500 $\mu$s and a peak power P$_{peak}$ of 1000 W is added to a P$_{CW}$ of 100 W. These results are simulated under the active fiber length of 2 m, pumping wavelength of 793 nm and R$_2$=0.5.}\label{sg1}
\end{figure*}

When we make the pump power evolve more slowly than the relaxation time of population from $^3H_4$ to $^3F_4$ by increasing the pump duration (500 $\mu$s), the temporal evolution of output power are illustrated in Fig. \ref{sg1}(a). In this case the temporal profile and duration of output power are the same with those of pump pulase, as seen in Fig. \ref{sg1}(a). The corresponding temporal behaviors of the population densities in the three energy levels are manifested in Fig. \ref{sg1}(b). The time span of population fluctuations nearly equals the duration of pump pulse. The population density N$_3$ of pump level temporally evolves in the same way as that of pump power, while the population densities N$_0$ and N$_1$ of lower and upper laser levels evolve in the opposite way as that of pump power as shown in Fig. \ref{sg1}(b).    

\begin{figure*}[htbp]
\centerline{\includegraphics[width=0.9\linewidth]{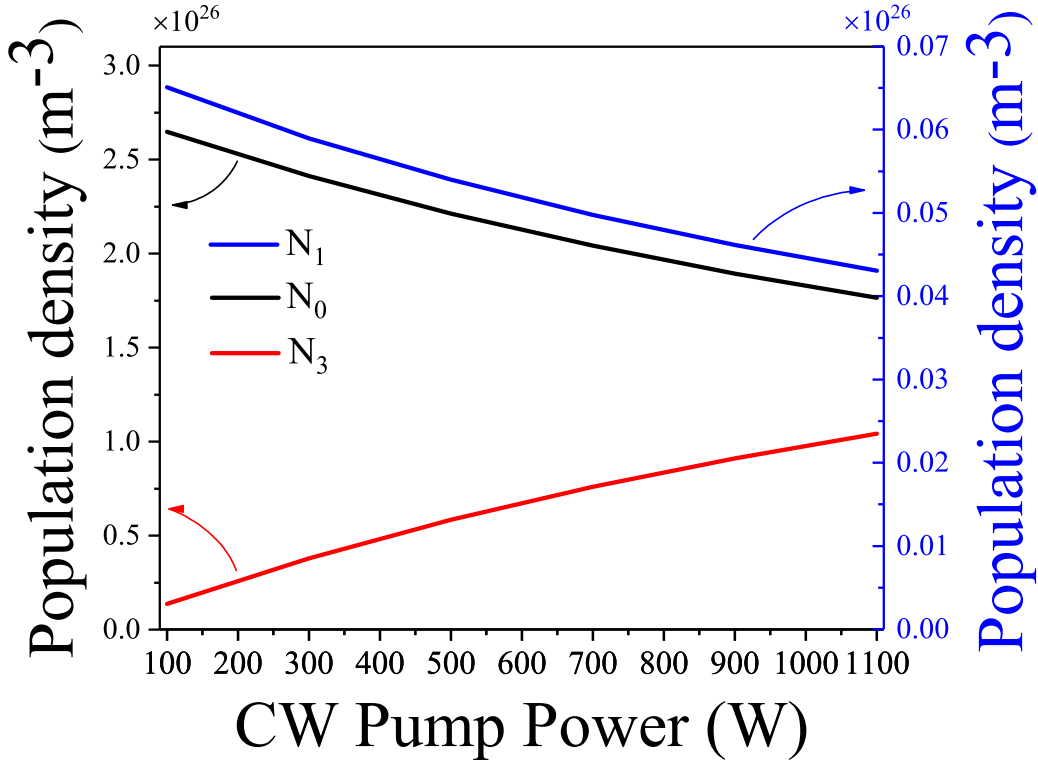}}
\caption{Dependence of the steady-state vaule of population densities (at the position of z=0) in the three energy levels on the CW pump power. These results are simulated under the active fiber length of 2 m, pumping wavelength of 793 nm and R$_2$=0.5.}\label{sg2}
\end{figure*}

We notice that the steady-state vaule of population densities reached by a CW power pumping is dependent on the CW pump power in $^3H_6$ $\rightarrow$ $^3H_4$ pump scheme due to the long relaxation time of population from pump level to upper laser level. The dependence of the steady-state vaules of the population densities in the three energy levels on the CW pump power is depicted in Fig. \ref{sg2}. The steady-state vaule of the population density in the pump level nearly increases linearly with the pump power, while the steady-state vaules of the population densities in the lower and upper laser levels nearly decrease linearly with the pump power, as displayed in Fig. \ref{sg2}. Comparing the values of the population densities under different pump powers in Fig. \ref{sg1} and Fig. \ref{sg2}, we find that during the pump power variation the population density of N$_1$(z=0,t$^\prime$) at time t$^\prime$ approximately equals the steady-state vaule of N$_1$(z=0) reached by a CW pumping with a power of P$_{CW}$+P$_{pulse}$(t$^\prime$), and this is also the case for N$_0$ and N$_3$. Consequently, the population densities in the three energy levels get clamped to their corresponding steady-state values under the pump power fluctuation created by a Gaussion pump pulse with a much longer duration than the lifetime of pump level. 

Since the temporal evolution of output power is identical with that of pump power and the population densities get clamped to their steady-state values, the stimulated emission and asorption are still balanced under the pump power flucatuation shown in Fig. \ref{sg1}. Therefore, pump power adiabatic evolution can be realized in $^3H_6$ $\rightarrow$ $^3H_4$ pump scheme provided that the duration of pump power is much longer than the lifetime of pump level. 

\subsection{Simulation results in $^3H_6$ $\rightarrow$ $^3H_5$ pump scheme}

\begin{figure*}[htbp]
\centerline{
\includegraphics[width=0.5\linewidth]{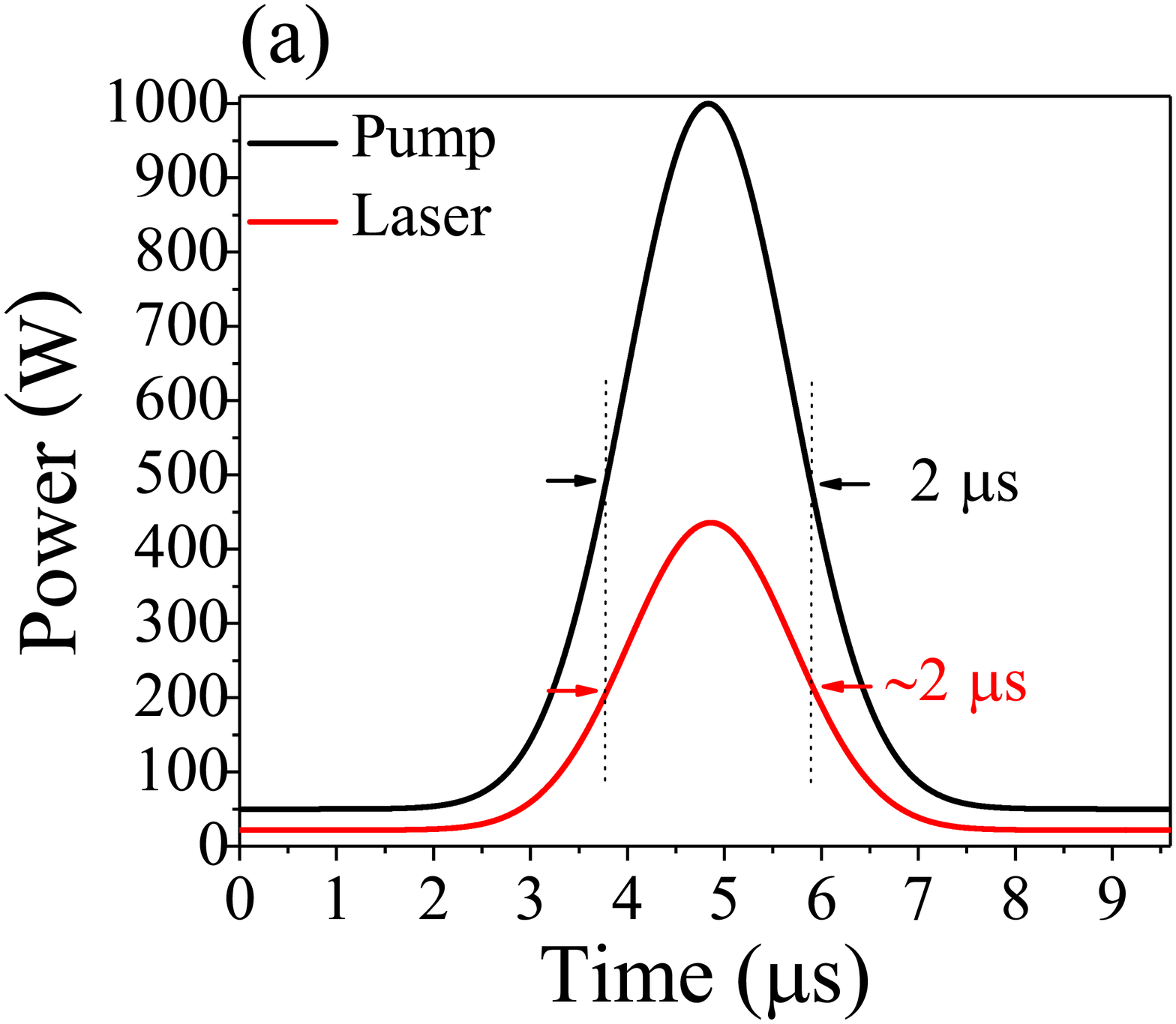}
\includegraphics[width=0.5\linewidth]{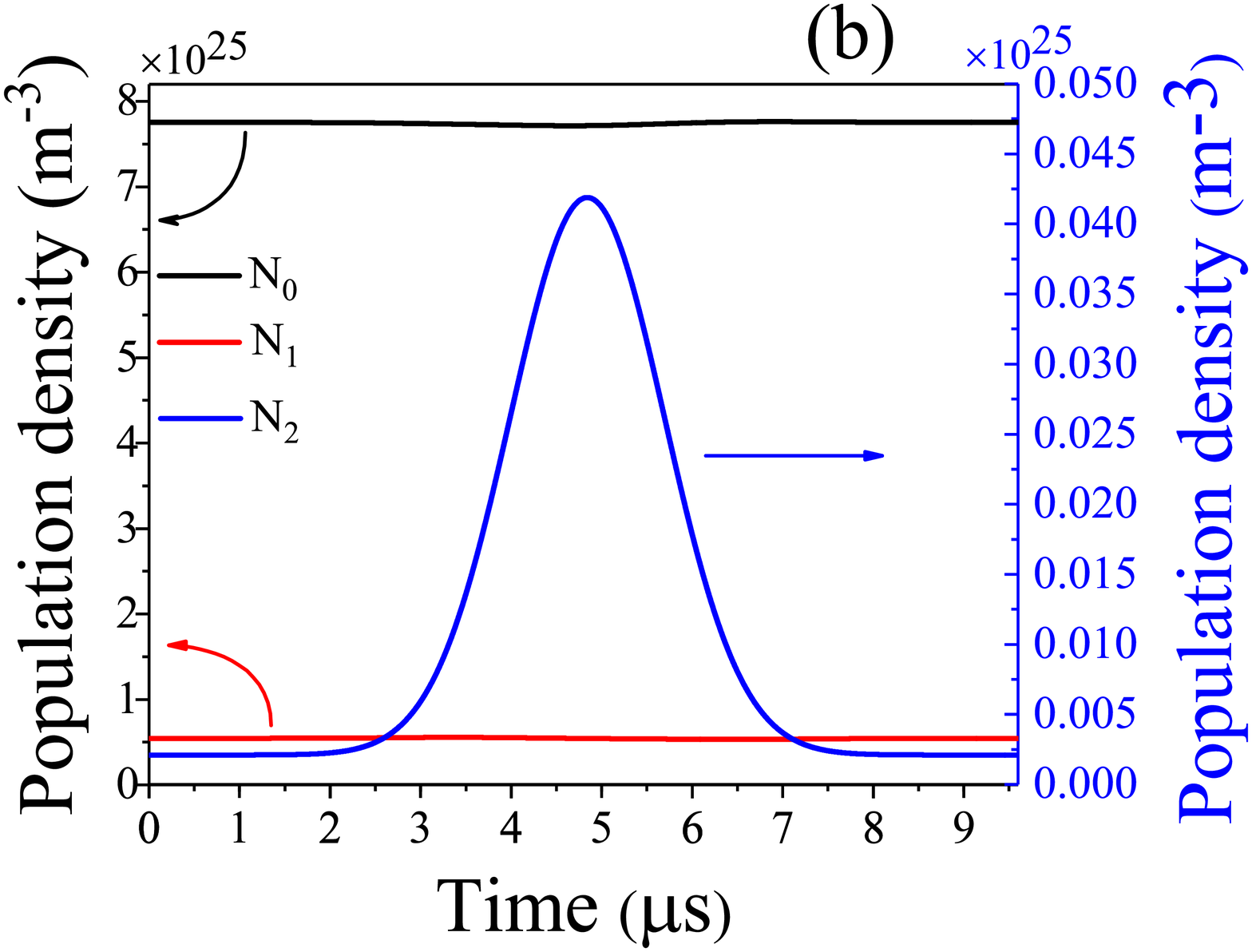}}
\caption{Simulation results without considering ESA: (a) The temporal evolution of the output power with that of pump, and (b) the temporal behaviors of N$_0$(z=0,t) in the lower laser level, N$_1$(z=0,t) in the upper laser level and N$_2$(z=0,t) in the pump level when a Gaussion pump pulse with a duration of 2 $\mu$s and a peak power P$_{peak}$ of 950 W is added to a 50 W CW pump power. These results are simulated under the active fiber length of 2 m, pumping wavelength of 1064 nm and R$_2$=0.1.}\label{sh}
\end{figure*}

In $^3H_6$ $\rightarrow$ $^3H_5$ pump scheme, we neglect the excited state absorption (ESA) and cross-relaxation processes, and investigate the laser behavior under pump power fluctuation in a model including the lower laser level $^3H_6$ (N$_0$), upper laser level $^3F_4$ (N$_1$) and pump level $^3H_5$ (N$_2$). In this case the temporal variation of laser power with that of pump is displayed in Fig \ref{sh}(a). A laser pulse is produced by the pump pulse P$_{pulse}$(t). As manifested in Fig \ref{sh}(a), both the duration and the temporal shape of the laser pulse are the same as those of pump pulse, which accords with the power evolutionary relationship between pump and laser in the power adiabatic evolution case.  

The temporal behaviors of population densities in the three energy levels are exhibited in Fig. \ref{sh}(b). Due to a very fast decay from the pump level (with lifetime of $\sim $7 ns) to the upper laser level, the population density N$_2$ of the pump level is much less than the population of N$_1$ (N$_0$) in upper (lower) laser level, so that we only be concerned with the populations N$_0$, and N$_1$. There is approximately no population fluctuation in the upper and lower laser levels under the pump power fluctuation, as manifested in Fig. \ref{sh}(b). Therefore, the stimulated emission and absorption are balanced as before under the pump power fluctuation, and a power adiabatic evolution of pump can be realized in $^3H_6$ $\rightarrow$ $^3H_5$ pump scheme without considering ESA.   

According to the simulation results, we conclude that the temporal evolutions of pump power, in $^3H_6$ $\rightarrow$ $^3H_5$, $^3H_6$ $\rightarrow$ $^3H_4$ and in-band pump schemes, can all become adiabatic when the duration of pump pulse, regardless of its peak power, is broader than a certain value called threshold duration. Here, the adiabatic evolution of pump power means the pump power evolves so slowly with time that the equilibrium between the stimulated emission and absorption is not changed. Analyzing the laser behaviors in the three pump schemes of Tm-doped fiber laser, we prove that the power adiabatic evolution of pump is an alternative pulsing technique, which can generate laser pulse with the same temporal shape and duration as those of pump pulse. Comparing with Q-switching and gain-switching, the duration of the laser pulse generated by the pulsing mechanism based on power adiabatic evolution of pump can be extended to a very long region. 

\section{Conclusion}
We have numerically demonstrated a CW operation of fiber lasers can tolerate a certain pump power upward fluctuation without destroying the balance between the stimulated emission and absorption. In this case the temporal variation of pump power is called power adiabatic evolution of pump, and it is proved to be an alternative pulsing mechanism in fiber lasers. To achieve adiabatic evolution the pump power should evolve slowly enough to ensure the duration of pump pulse is broader than a certain value called threshold duration. The threshold duration can be governed by adjusting the CW pump power, reflectivity of cavity and the temporal profile of pump pulse. Since the stimulated emission and the absorption still compensate one another under power adiabatic evolution of pump, the laser pulse generated by this mechanism has the same duration and temporal shape as those of pump pulse. The peak power of laser pulse can also be controllable by managing that of pump pulse and the optical efficiency. The adiabatic condition permits us to employ a pump pulse with a long duration and a high peak power, which can achieve a laser pulse generation with a long duration and meanwhile a high peak power. Those are the main advantages of pulsing mechanism based on the power adiabatic evolution of pump compared with Q-switching and gain-switching methods.

\end{document}